\documentclass[preprint,prb,amsfonts,amsmath,amssymb]{revtex4}
\usepackage{bbding}
\usepackage{pifont}
\usepackage{graphicx}
\usepackage{amssymb}
\usepackage{txfonts}
\usepackage{dcolumn}
\usepackage{amsmath}
\usepackage{bm}

\usepackage{colordvi}

\begin{document}
\title{Gibbs Paradox in the View of Information Entropy}
\author{Xiao XU}
\email{xuxiao@scut.edu.cn}
\affiliation{School of Physics and Optoelectronics, South China University of Technology, Guangzhou 510651, China }


\begin{abstract}
This paper introduces the basic concepts of information theory. Based on these concepts, we regard the states in the state space and the types of ideal gases as the symbols in a symbol set to calculate the mixing entropy of ideal gas involved in Gibbs Paradox. The discussion above reveals that the 'non-need for distinguishing' can resolve the contradiction of Gibbs Paradox, implying the introduction of ‘indistinguishability’ is not necessary. Further analysis shows that the information entropy of gas molecular types does not directly correlate to the energy of a gas system,so it should not be used for calculating thermodynamic and statistical dynamic entropies. Therefore, the mixing entropy of the ideal gas is independent of the molecular types and is much smaller than the value commonly thought.

\end{abstract}

\maketitle

In Shannon’s article\cite{1949Shannon}, the concept of information entropy stems from sources that are completely different from thermodynamics and statistical mechanics, and it also does not refer to statistical methods of statistical mechanics as physicists usually think\cite{2020CollPhy}. It is said that Von. Neumann noted the similarities between Shannon's formula and Boltzmann's entropy formula and thus he suggested Shannon named the average information quantity as "entropy ".\cite{2003Avery}  Actually, the concept of information entropy is mainly established by engineers and mathematicians, which is applied in a very specific field and its development is very straight forward, unlike that of thermodynamics and statistical mechanics which involves a series of philosophical discussions and debates. Therefore, the theory of information entropy, as a mathematical tool, provides a "filter" to examine the thoughts and experiments of thermodynamics and statistical mechanics.

This paper will introduce the concepts of and about information entropy; and based on these concepts, we will expound and prove the following two statements :(1) If we analyze thermodynamic entropy based on information theory, the contradiction of Gibbs Paradox will be dissolved, and the mixing entropy of the same particle will vanish as well. Therefore, the conception "indistinguishability" is not necessary. (2) The concern of informational entropy is coding, not energy conversion. " Information "and" energy " have no essential connection. When this feature is emphasized, it is found that there is no mixing entropy increase directly corresponding to the energy relationship of different types of particles.

\section{Basic Concept of Information Theory}

Since Morse had invented Morse code which uses dots and dashes as binary symbols to express information, people have begun to use binary in communication. As we all know, a basic digit of binary is called a “bit”, named by J.W.Tukey\cite{1949Shannon},\cite{2006Bit},which is the most used unit in information theory.

The most important purpose of information theory is to determine the shortest length of the encoded bit stream when we encode a symbolic sequence-such as an English telegram consisting of the letters, spaces and punctuation marks-in telecommunication.

From this starting point, we will introduce in this section the concepts of information quantity and information entropy, conditional entropy and average mutual information, processing of continuous signals, typical sequences, in order to use these concepts to analyze problems related to thermodynamics and statistical mechanics.

\subsection{Information Quantity and Information Entropy}

If there are four symbols to be coded in binary codes, the simplest way is to encode them with 00,01,10,11, so we have 2 bits of encoding. This bit number can be calculated by the formula ${\rm log}_{2}4$ . The formula,${\rm log}_{m}n$ , is used for calculating the number of code digits, where $m$ is the number of symbols in the source alphabet, and $n$ is the base of the numeral system (i.e. the number of symbols in the code alphabet). The earliest meaning of information quantity is exactly the number of the code digits, which was put forward by Hartley in 1928.\cite{1928Hartley}

However, this coding approach is not able to use encoding resources efficiently enough, because, in practical use, some source symbols have higher probabilities of appearing, while some symbols have lower ones, and the aforementioned coding method does not take probability into account. A more reasonable approach should encode symbols with higher probabilities into fewer digits while symbols with lower probabilities into longer digits. In such case, the number of code digits of a symbol , i.e. the information quantity, should be ,$${\rm log}_{m}\frac{1}{Pr(x_{i})},$$ where$Pr(x_{i})$ is the appearing probability of $x_{i}$.

The average number of code digits of each symbol over the whole system, which is named “entropy” by Shannon, can be calculated as,
       \begin{equation}
H(X)=-\sum_{i}Pr(x_{i}){\rm log}_{m}Pr(x_{i})
 \end{equation}
where '$H$'of $H(X)$, which is borrowed from Boltzmann's H theorem, represents the information entropy of the information source with the symbol set $X$.\cite{1949Shannon}

\subsection{Conditional Entropy and Average Mutual Information}

The purpose of communication is to transmit information. Shannon once said: " The fundamental problem of communication is that of reproducing at one point either exactly or approximately a message selected at another point."\cite{1949Shannon} But channels for transmitting signals have noise and interference that can lead to distortion of signals. A concrete embodiment of the distortion is that the voltage or current signal of a certain symbol from the source changes the waveform because of noise or interference in the channel, and thus the signal is decided to be another symbol at the destination. Of course, this kind of misjudgment does not always occur, but happens randomly at a certain probability of.

In order to deal with this kind of noise or interference, in telecommunication, error detecting and correcting codes are added to the original codes from the source to detect and correct the error of signals at receiver. Then how many digits for source coding should be used in ensuring of with good efficiency and how many digits for transmitting with good error detection and correction as well? Information theory uses abstract theory to evaluate the optimal code digits and brings two concepts: conditional entropy and average mutual information.

In order to make these two concepts well understood, information theory introduces another concept: uncertainty.

When a source does not send a message, we do not know what message it will send next moment. As a result, the next sent message is uncertain for us. Obviously, the more code digits a source needs, the greater the entropy of information, the greater the uncertainty to us. Once we receive a symbol at the next moment, the uncertainty disappears, which eliminates the uncertainty corresponding to information quantity belong to the symbol.

However, due to interference, the symbol we get at the receiver may be not the corresponding symbol, so this uncertainty cannot be completely eliminated. Therefore, conditional entropy can be defined to measure this uncertainty, as:
     \begin{equation} H(X/Y)=-\sum_{i,j}Pr(x_{i}y_{j}){\rm log}_{m}Pr(x_{i}|y_{j})\end{equation}
where $X$, the sender, is regarding to issuing symbol $x_{i}$, where the receiver, $Y$, receiving the symbol $y_{j}$; $H (X/Y)$ can be seen as the average measure of uncertainty of the transmitting, that is , the average digits of information lost due to channel interference.

It is now easy to introduce the concept of average mutual information:
    	   \begin{equation}I(X;Y)=H(X)-H(X/Y)\end{equation}
$I (X; Y)$ represents the average information quantity transmitted through the channel after overcoming interference.

Based on these two concepts, we can make a simple and intuitive result on how to code for transmitting: We need $H (X / Y)$ digits-- as detecting and correcting digits-- to fix the damage caused by interference.

Additional knowledge is needed to figure out how to encode. In this article, we just introduce the most basic concepts of information theory. Hence, we do not introduce encoding method here.

\subsection{Processing of Continuous Signals}

 In the field of communication, we always try to address continuous waveform signals, for example, voltage or current waveform signals that represent sound through microphones and recording devices. Such signals should have information quantities before and after channel transmission. Because the waveform signal is continuous in the time domain and the amplitude domain, in order to be processed according to the above idea, the signal must be discretized in the time and the amplitude domain respectively.

If the signal has the highest frequency component, we can sample the signal with the rate which is more than twice the highest frequency by using the method specified in the Nyquist sampling theorem. After this sampling, we can obtain a time-discrete series signal without information lost. The signal is discretized in the time domain. Because this discrete process is not particularly helpful for us to understand the contents in this paper, we will not go into details. If interested, readers can refer to the relevant content in the course textbooks for Signal and Systems.\cite{2001Oppenheim}

In the amplitude domain, we quantize the continuous amplitude $x\in [a,b]$ . Uniform quantization is used here. When the real number interval  is divided into $n$ parts, then each quantizing step size is$\Delta x=(b-a)/n$ , and the $i$th interval is $[i \Delta x+a,(i+1)\Delta x+a]$($i=1,2,...,n-1$). Then if the probability density function of the random variable $X$ is $f_{X}(x)$ , the probability that the value of the $X$ belongs to the $i$th interval is
   \begin{equation} \int_{i \Delta x+a}^{(i+1)\Delta x+a}\approx f_{X}(\frac{(2i+1)\Delta x}{2}+a)\Delta x\end{equation}

After discretization in this way, the entropy of this random variable can be defined:
         
    \begin{equation} H(X)=-\sum_{i=0}^{n-1}((\frac{(2i+1)\Delta x}{2}+a)\Delta x){\rm log}_{m}((\frac{(2i+1)\Delta x}{2}+a)\Delta x)\end{equation}
When $\Delta x$ is small,
 \begin{equation} H(X)\approx-\int_{a}^{b}f_{X}(x){\rm log}f_{X}(x)dx-{\rm log}_{m}\Delta x\end{equation}
Let $n\rightarrow\infty,\Delta x\to 0$, then
    \begin{equation}
H(X)=-\int_{a}^{b}f_{X}(x){\rm log}f_{X}(x)dx-\lim_{\Delta \to 0}{\rm log}_{m}\Delta x
\end{equation}
Clearly, as ,$H (x)$ tends to infinity. This is in line with our expectation that a continuous signal, to be finely characterized, has an infinite number of states and will also bring infinite information quantity. Such a definition seems to have no practical significance. However, considering in real cases, the evaluation of continuous signal transmission is relative to noise. If the same quantization criteria are used for noise and signal, the number of code digits of noise and signal increases synchronously, while the difference of relative code digits does not change. Therefore, the \emph{continuous entropy} is defined as follows:
   \begin{equation}  
H_{c}(X)=-\int_{a}^{b}f_{X}(x){\rm log}f_{X}(x)dx
\end{equation}

In communication, continuous entropy can reflect the difference of amounts of information between signal and noise.

The reason why we introduce continuous entropy here is to prepare for the later analysis of the mixing entropy of different kinds of particles in Gibbs Paradox.

\subsection{Typical Sequence}

Typical sequence is the content of asymptotic analysis in probability theory, which is not a unique part belonging to information theory.\cite{2006Cover} In the field of physical research and education, its ideas are often used, such as the approach of thermodynamic limits, the most probable distribution, etc. But often there is no relevant knowledge to be explained. Here is a brief introduction.

Take the production of a binary sequence as an example to explain what a typical sequence is. If the symbol ‘1’ and ‘0’ occur equal likely, and the symbols in the sequence are statistically independent, then the sequence appears as follows:

0011011011100100...... is a \emph{typical sequence}, with the frequency of 50\% for 0 and for 1 respectively.
Conversely, if a sequence occurs, such as:
1111101110001111... is an \emph{atypical sequence} with a frequency of 75\% for 1 and 25\% for 0.\\

To sum up, if the frequency of that each character in the sequence is equal to its probability, it is a typical sequence; conversely, if not equal, it is an atypical sequence. 

\emph{As the sequence length increases, the sum of the probabilities of all the typical sequences approaches 1 with each typical sequence occurring equally likely. Thus the sum of probabilities of all the atypical sequences trends to 0. }This feature for random sequences is called asymptotic equipartition property.

We will use this point when we discuss Boltzmann entropy later. (An aforementioned sequence is characterized by the statistical independence of each character. In the case that each character is not statistically independent, if the sequence is stationary, it can be processed with the similar idea as typical and atypical sequences.) 

\section{The Treatment of Gibbs Paradox}
The Gibbs paradox is explained here.\cite{2012Pathria}

   \begin{figure}[ht]
\includegraphics[scale=0.6]{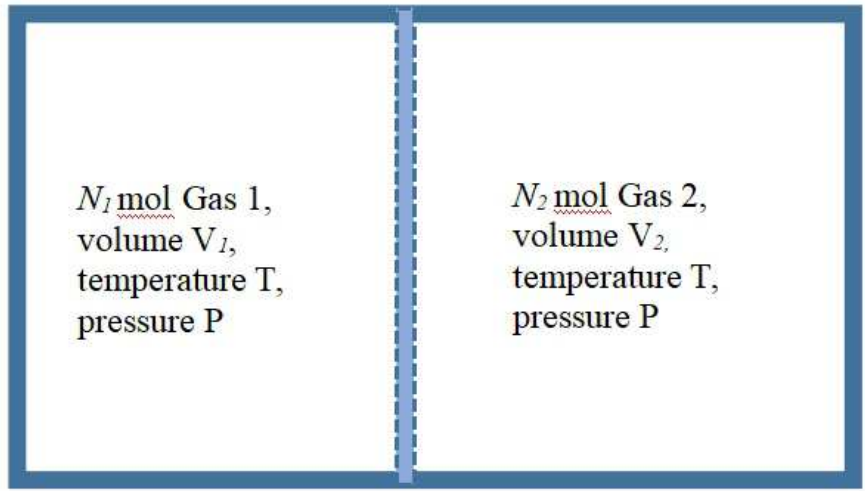}
\caption[Figure.1] {The Illustration of Gibbs Paradox}
\end{figure}                       

As shown in figure 1, the volume space, $V_{1}$ and $V_{2}$ are separated by a board. There are two different ideal gases, Gas 1 and Gas 2, each with $N_{1}$ and $N_{2}$ particles are under pressure $P$ and temperature $T$. When the board is removed, the two gases diffuse to each other's space, and when the two gases are filled with the whole volume space $V=V_{1}+V_{2}$ , the entropy of the system in the whole cavity increases as:
 \begin{equation}
\Delta S=(N_{1}+N_{2})k_{B}{\rm ln}V-N_{1}k_{B}{\rm ln}V_{1}-N_{2}k_{B}{\rm ln}V_{2}
\end{equation}

It is easy to understand that after removing the board, the system neither does external work nor absorbs heat from the outside, so the work of the system to the outside and $Q/T$ are zero. However, because the diffusion process of gases is non-reversible, according to Clausius' theorem,$Q/T<\Delta S$ ,entropy increases,$\Delta S>0$. This increased entropy is called the mixing entropy of gases.
In particular, when:$V_{1}=V_{2}=V/2,\qquad N_{1}=N_{2}=N/2$,
   \begin{equation}           \Delta S=Nk_{B}{\rm ln}2 \end{equation}

Now, if we replace Gas 1 and Gas 2 with the same gas, all the gas molecules must diffuse to the whole cavity, so it is an irreversible process, and there are still mixing entropy as (9) and (10).

However, for the same gas molecules, this diffusion process will not bring any changes in the macro system, and the chemical potential of the gas will not change, so there should be no entropy increase and the mixing entropy should be 0. "The mixing entropy is 0" contradicts with the formulae (9) and (10), and the contradiction produces a paradox.

\subsection{Background of the Emergence of the Gibbs Paradox}

In 1850, after Joule's work overturned the calorie theory, Clausius retained Carlo's law. The entropy of the system in equilibrium state was defined, and Clausius's law was given:$dQ/T\le dS$. This law shows that for systems in which the initial and final states are equilibrium states, if passing through the irreversible paths, the system endothermic ratio is less than that though the reversible path; in other words, for a non-equilibrium process, the entropy of the system changes more than that through an equilibrium one. This part of the entropy produced by the change of the system itself without heat exchange is due to the spontaneous trend of the system: a system always tends to the maximum entropy under the constraint conditions, and reaches the thermodynamic equilibrium state.\cite{1879Clausius} That is to say, when equilibrium, the entropy is the maximum under the constraint condition and the free energy is the lowest. It is an important criterion for the equilibrium or generation of chemical reaction. Since 1872, Gibbs has used Clausius's theorem to create his theory of phase transition and phase diagram,\cite{1875Gibbs},\cite{1877Gibbs}and these theories were recommended by European chemists and entered the Mainland of Europe at that time to solve the long-standing chemical equilibrium determination problems of chemists.cite{1910Jodan}

The Gibbs Paradox is found in ‘the Fundamental equations of Ideal Gases and Gas Mixtures section of the article On the Equilibrium of the Heterogeneous Substances’ in 1875.\cite{1875Gibbs},\cite{1877Gibbs} Its derivation and proposal are not based on gas molecular motion theory and statistical mechanics, but on thermodynamics. This article, which is to solve the multiphase equilibrium, gas diffusion must be considered.

Gas diffusion is an irreversible process, and it is necessary to increase the corresponding entropy, which is the subtext after Gibbs sets this paradox.\\
\subsection{Consistency between information entropy and Boltzmann entropy}
In order to use the concept of information entropy, we first check the consistency between the Boltzmann entropy and the information entropy. Obviously, if there is no consistency, then we cannot use information entropy to discuss statistical mechanical entropy. The check begins with the Boltzmann entropy because Gibbs began thinking about the definition of entropy from a statistical mechanics point of view in 1902, from the Boltzmann entropy, that is,\cite{1902Gibbs}
           \begin{equation}      S=k_{B}{\rm ln}W \end{equation}
where $W$ refers to the number of microscopic states of the system. What is the number of microscopic states and how to calculate the number of them involves two problems: one is which physical quantities are selected as parameters to describe the state of a particle; the other is how the parameters are discretized if they are continuously distributed quantities, such as the momenta, the position coordinates, etc.

In retrospect, Gibbs chose a point in the phase space of a particle to describe the state of the particle, so the momentum and the geometric position became selected parameters; in addition, the inner freedom degrees of the particle should be considered; and for the interactions between particles -they are what chemists had to face and Gibbs had to consider-should be considered too. Obviously, the interactions between different kinds of particles may be different from the same kind of particles, so the type of particle needs to be defined as a parameter.

Among these parameters,the positions and momenta are continuous.Hence,in this paper, the momentum coordinates and the geometric space position coordinates are uniformly quantized, and then let these quantizing step size tend to infinitesimal (as shown in formulas (8),(9)). Finally, the processing of continuous variables can be consistent with the continuous entropy.

In classical statistical physics, the continuous variables can be processed straightforwardly as described above. Then we must quantize the interval of momentum $\Delta p$ and the interval of geometric position $\Delta q$ in a certain way what makes each continuous parameter of the system be uniformly quantized and finally discretized. The informational entropy can be calculated under the discretized parameter condition;futher,it's value should be the sumation of the continuous entropy and a term including ${\rm ln}\Delta p \Delta q$.

In quantum statistics, the size of the phase cell is fixed in the form of a power of $h$, the Planck constant, that is, the size of is always $h$. However, the fixed one is $\Delta p \Delta q$ , not $\Delta p$or$\Delta q$. Hence, it does not affect the quantization process with handling of the Gibbs Paradox, as described in the later section on quantum mechanics.

$k_{B}$, the Boltzmann's constant, is simply a conversion to the dimension $[J K_{-1}]$, will not affect the discussion of the consistency of the two entropy.\\
Now focus on the problem${\rm ln}W$ . If the particles in the system are huge and the interaction of particles is mainly limited to the neighbors in their geometric position space, then the theory of the previous typical sequence shows that only those states that accord with the most probable distribution of particles accounts for the vast majority of states (of the system) and occur equally likely. As a result, a state of the system corresponds to a symbol in information theory, and when the number of the particles of the system is huge, the corresponding information entropy of the system is:

\begin{eqnarray}
\lefteqn{H =  -\sum_{i=1}^{n} Pr(i)\ln Pr(i){}}
\nonumber\\
& & {}\xrightarrow{n\to \infty}\sum_{i=1}^{n}\frac{1}{W}\ln W=\ln W
\end{eqnarray}  
In other words, when the state space is selected and a state of the state space is presented as a symbol, the definition of Boltzmann entropy and the definition of information entropy are compatible for systems with large particles. Although those states that corresponding to atypical sequences are also calculated in Boltzmann's formula, finally they are still ignored because of the use of Stirling formulas.

\subsection {Information Entropy Processes the Gibbs Paradox}

In the Gibbs Paradox, for each gas molecule, the scale size of each dimension of the phase space is divided at first, and the size of each phase cell is $h^{3}$. The center coordinates of the quantized cell of a momentum subspace are $\boldsymbol{p_{i}}$ , and the cell size is $\Delta V_{p}$;the center coordinates of the quantized cell of the position subspace are $\boldsymbol{r_{j}}$, and the cell size is $\Delta V$;the types of the gas molecules are expressed as $\mu_{k}$,and there are two kinds of gas molecules in this paradox, $\mu_{1}$ and $\mu_{2}$. The symbol set is defined as $\{(\boldsymbol{p_{i}},\boldsymbol{r_{j}},\mu_{k})\}$, which means that the particle is in the $i$th momentum cell and the $j$th position cell, and its type is  $\mu_{k}$.

For simple presentation, we only deal with condition $V_{1}1=V_{2}=V/2,N_{1}=N_{2}=N/2$ here.For $V_{1}$,the subscript of $\boldsymbol{r_{j}}$, $j$ is from 1 to $V/(2\Delta V)$;for $V_2$, the $j$ is from $V/(2\Delta V)+1$ to $V/(\Delta V)$.

For different particles, let the initial particle type in $V_{1}$ is $\mu_{1}$,and the type in $V_{2}$ is $\mu_{2}$.For $V_{1}$, when the board is not taken away, the information entropy of a particle inside is
\begin{eqnarray}
\lefteqn{H_{s_{1}} =  -\sum_{i,j;\boldsymbol{r_{j}}\in V_{1}} Pr(\boldsymbol{p_{i}},\boldsymbol{r_{j}},\mu_{1})\ln Pr(\boldsymbol{p_{i}},\boldsymbol{r_{j}},\mu_{1}){}}
\nonumber\\
& & {}=-\sum_{i}Pr(\boldsymbol{p_{i}})\ln Pr(\boldsymbol{p_{i}})-\sum_{j=1}^{V/(2\Delta V)}Pr(\boldsymbol{r_{j}})\ln Pr(\boldsymbol{r_{j}})
\nonumber\\
& & {}=-\sum_{i}Pr(\boldsymbol{p_{i}})\ln Pr(\boldsymbol{p_{i}})+\ln V-\ln (2\Delta V)
\end{eqnarray}                              
In the above derivation, the independence of $\boldsymbol{p_{i}}$ and $\boldsymbol{r_{j}}$ in the classical statistical mechanics are used,  and the probability of  occupying the cell of the same volume size is the same, i.e.$Pr(\boldsymbol{r_{j}})=2\Delta V/V$ .

By the same token, the information entropy of a particle in $V_{2}$  is ,
\begin{equation}
H_{s_{2}} = -\sum_{i}Pr(\boldsymbol{p_{i}})\ln Pr(\boldsymbol{p_{i}})+\ln V-\ln (2\Delta V)
\end{equation}   
           
After removing the board, a particle's optional set of symbols includes all the possible momenta, positions, and types, so its informatin entropy is
      \begin{eqnarray}
      \lefteqn{H_{D}=-\sum_{i,j,k}Pr(\boldsymbol{p_{i}},\boldsymbol{r_{j}},\mu_{k}) \ln Pr(\boldsymbol{p_{i}},\boldsymbol{r_{j}},\mu_{k}){}}
\nonumber\\
&&{}=-\sum_{i}Pr(\boldsymbol{p_{i}}) \ln Pr(\boldsymbol{p_{i}})-\sum_{j=1}^{V/\Delta V}Pr(\boldsymbol{r_{j}}) \ln Pr(\boldsymbol{r_{j}}) -\sum_{k=1}^{2}Pr(\mu_{k}) \ln Pr(\mu_{k})
\nonumber\\
&&{}=-\sum_{i}Pr(\boldsymbol{p_{i}}) \ln Pr(\boldsymbol{p_{i}})+\ln V -\ln \Delta V+\ln 2
\end{eqnarray}      
where the calculation of $Pr({\mu_{k}})$ is based on the condition that the number of particles of different kinds is equal, i.e.$Pr(\mu_{1})=Pr(\mu_{2})=1/2$.

In the case of large number of particles,$Pr(\boldsymbol{p_{i}})$ are the same in $H_{S_1}$, $H_{S_2}$ and $H_D$, as the momentum states of the system has almost no change before and after drawing the board. Combined with (13),(14), the information entropy difference of the system before and after drawing the board is:
 \begin{equation}
          \Delta H=N H_D-\frac{N}{2}H_{S_1}-\frac{N}{2}H_{S_2}=N(2 \ln 2)
\end{equation}
By comparing Formula (16) with Formula (10), we will find that--apart from the coefficient,$k_B$, related to energy conversion--the value of information entropy is twice as high as the value of the Gibbs entropy we usually regard!

If there are the same particles in the two regions, as particles $\mu_1$,then $Pr(\mu_1)=1$ . In this kind of system, the entropy of a particle after drawing the board is:
         \begin{eqnarray}
      \lefteqn{H_{D\_ same}=-\sum_{i,j,k}Pr(\boldsymbol{p_{i}},\boldsymbol{r_{j}},\mu_{k}) \ln Pr(\boldsymbol{p_{i}},\boldsymbol{r_{j}},\mu_{k}){}}
\nonumber\\
&&{}=-\sum_{i}Pr(\boldsymbol{p_{i}}) \ln Pr(\boldsymbol{p_{i}})-\sum_{j=1}^{V/\Delta V}Pr(\boldsymbol{r_{j}}) \ln Pr(\boldsymbol{r_{j}}) -\sum_{k=1}^{1}Pr(\mu_{k}) \ln Pr(\mu_{k})
\nonumber\\
&&{}=-\sum_{i}Pr(\boldsymbol{p_{i}}) \ln Pr(\boldsymbol{p_{i}})+\ln V -\ln \Delta V
\end{eqnarray} 
then:
 \begin{equation}
          \Delta H_{D\_same}=N H_{D\_same}-\frac{N}{2}H_{S_1}-\frac{N}{2}H_{S_2}=N( \ln 2)
\end{equation} 
In (16) there are two ln2.The origins of the two $\ln2$ are different. One is from the uncertainty of the particle in different positions, the other is from the uncertainty of the particle type. The two $\ln2$ have the same value, but it's just a coincidence: because the different particle number ratio is the same with volume ratio $V_1/V_2$, so the entropy related to the position is exactly the same as the entropy related to the particle type. Hence, in Formula (18), the entropy introduced by different particle types is reduced.

The Gibbs Paradox is not really a paradox, just because we confused the entropy related to particle types with the entropy related to positions. Gibbs thought that the entropy of mixing was the mixture of different particles, so it was mainly related to the type of particles, not to the positions.

In the following analysis, I will prove that when the number of particles is huge, the entropy increase related to the positions will be offset by the conditional entropy brought by the " non-need for distinguishing" property. In this way, the Gibbs Paradox is dispelled.

Here, I define a new noun called particle's " non-need for distinguishing" property. Because the thermodynamic and chemical properties of the system do not be determined by which particle is in which state, but how many particles are in which state, it is unnecessary to distinguish each particle’s concrete position when considering the calculation of thermodynamic properties of the system. 

Now, to make the " non-need for distinguishing" property easy to understand , I use an example of information theory to illustrate as following.

\leftline{$\blacktriangleright$}
Example 1. For example, there is a string sequence of length 3, of which each position can take one of the four character {a,b,c,d} equally likely and statistically independently. A set $X$ consisting of these sequence has $4^3=64$ elements, and the probability of occurrence of one certain string is $Pr(x_i)=\frac{1}{64}$ .Regarding $X$ as an information source ,the entropy of the source is,
$$H(X)=-\ln(1/4^3)(Nat)\approx4.16Nat$$
or
$$H(X)=-\log_{2}(1/4^3)(Bit)\approx6Bit$$.

Considering strings $\{abc; bca; cab; acb; cba; bac\}$, all these strings are "one a, one b and one c ". If we only focus on "one a, one b and one c " without paying attention to character’s order, we get an new information source, Source $Y$, which entropy can calculated by conditional entropy $H (X / Y)$ shown as Formula (2). (Readers interested in computing pay attention here: Like for the sequence “acb”, we can get each $Pr(x_i|y_j)$. At last we have $H (X/Y)=1.68(Nat)=2.42(Bits)$. )

By using the Formula (3) for calculating average mutual information, we can get:
$$I(X;Y)=H(X)-H(X/Y)=3.58(Bit)$$

Clearly, if the number of optionable characters is very large, far greater than the number of positions n, we get,
$$H(X/Y)\approx \ln (n!)$$
$$I(X;Y)=H(X)-\ln(n!)$$
                                                \rightline{$\blacktriangleleft$}\\

We compare the concrete positions of each particle as the position of each character of the string in Example 1, then removing the information of the concrete positions of the particle, we can calculate the entropy by I(X;Y) as the example. In general, the number of the particle's states are far greater than the number of particles,$I(X;Y)=H(X)-\ln(n!)$ can be used for calculating the entropy of a particle considering the " non-need for distinguishing" property. (In the following part, to make the derivation easy to understand, we use '$H_I$ ', but not '$I$ ', to denote average mutual information). Referring to from the Formula (13) to (16), we have:
 \begin{equation}
          \Delta H_D=N H_D-\ln(N!)-\frac{N}{2}H_{S_1}+\ln((N/2)!)-\frac{N}{2}H_{S_2}+\ln((N/2)!)
\end{equation} 
If the Stirling formula is used as a first approximation, then:
 \begin{equation}
-\ln(N!)+2\ln((N/2)!)\approx -N\ln N+N+N(\ln N-\ln 2)-N=-N ln2
\end{equation} 
Replace (20) with (19), refer to (16), we have:
\begin{equation}\Delta H_I\approx N ln2\end{equation}			

For the same particles in two regions, refer to formula (18), similarly, we have:
                \begin{equation}\Delta H_{I\_same}\approx 0\end{equation} 

\subsection{ "Non-need for distinguishing" and " Indistinguishability"}

Although a lot of researchers have realized that the indistinguishable nature of quantum mechanics is not necessary to explain the removal of particle positions’ arrangement in calculating the entropy, their descriptions are mostly vague. So here, I give an easy-to-understand explanation.

As shown in figure 2, suppose there are two particles whose states correspond to coordinate $X_1$ and $X_2$ respectively. If the coordinates are quantized and separated, the combined states of the two particles can be $(i ,j)$, as shown in Figure 2.

  "Indistinguishability", introduced by Bose\cite{1924Bose}, means that State $(1,3)$ and $(3,1)$ are actually the identical state despite of which particle in State 1 or in State 3. Therefore, in the microcanonical ensemble, if we still use the assumption that the probability of occurrence of various states is the same, and for convenience, we still retain the notation of $(1,3)$ and $(3,1)$, but these two states are the same state, then there are:
    \begin{equation}Pr(1,1)=2Pr(3,1)=2Pr(1,3)\end{equation} 
That is, for two particles--each particle has only one state--the probability of a diagonal element is twice the probability of a non-diagonal element.

  \begin{figure}[ht]
\includegraphics[scale=0.6]{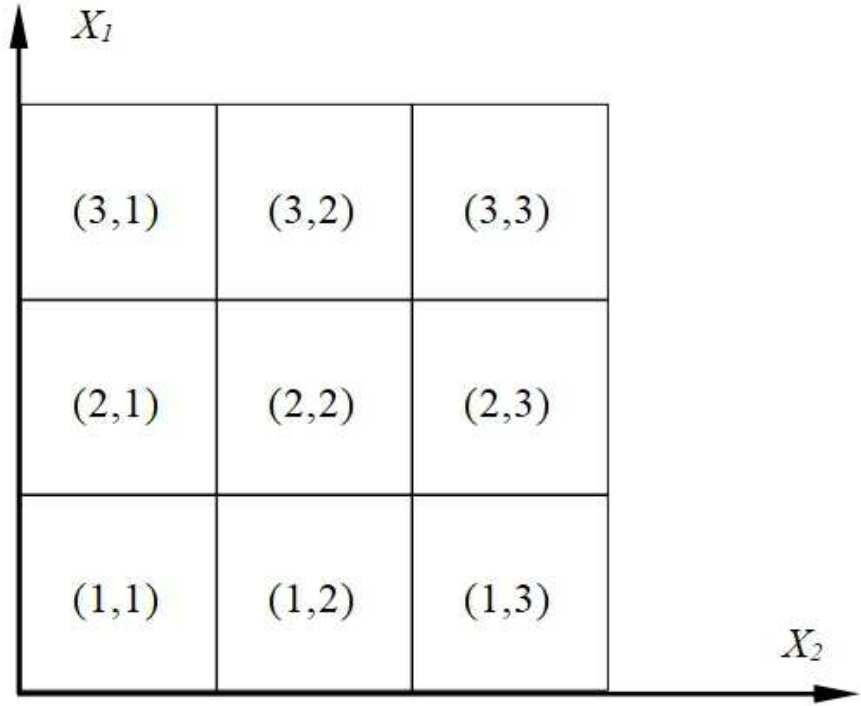}
\caption[Figure.2] {Particle State Diagram}
\end{figure}                             

The so-called “non-need for distinguishing” is that we still retain Gibbs' hypothesis when he established the ensemble theory, and each cell probability is the same, that is:
    \begin{equation}Pr(1,1)=Pr(3,1)=Pr(1,3)\end{equation}
And we only convert $(3,1)$ and $(1,3)$ into one case without considering the particle’s arrangement.

With the increase of states, it is obvious that the proportion of diagonal elements in the whole system is getting lower and lower, so the probability of being taken is getting lower and lower. Therefore, there is no difference between “non-need for distinguishing” and "indistinguishable ", and the calculations of entropy from two perspectives are exactly the same.

When the number of states or of particles are not big enough, it cannot be ignored the result of drawing the board for a particle moving in a greater space and bringing about entropy increase. This ignorance has revealed twice in the previous calculation process. One is the use of  in formula (19) and the other is the use of Stirling formula in formula (20), respectively.

\section{Treatment of the problem of different types of particles}

For scholars studying the Gibbs Paradox, the real problem of the Gibbs Paradox is not that we have long focused on the mixing of the same molecules in thermal textbooks, but that different molecules. No matter how different those molecules, as long as they are different, they will bring a entropy increase. This conclusion is not “physical”.\cite{1992Jaynes} Because $k_B$ is related to energy, so the mixing entropy should reflect the characteristics of the energy. And real particles of different kinds should have different energy properties, such as mass, internal degrees of freedom, etc., cannot all be the same entropy increase. Furthermore, even the ideal gases have different particle masses. To this problem, Gibbs himself interprets it as an imperfect ideal gas model.\cite{1875Gibbs}

Now, I use new reasoning to show that different kinds of ideal gases do not have thermodynamic and statistical mixing entropy. 

It is known that the entropy of an ideal gas is\cite{2012Pathria}:
 \begin{equation}S_i=N_i k_B \ln V+\frac{3}{2}N_i k_B \ln \bigg(\frac{2\pi e m_i k_B T}{h^2}\bigg)\end{equation}
where, $N_i$, the number of ideal gas molecules of Type $i$; $m_i$, the mass of a gas molecule of Type $i$. Now, the information entropy of the molecules of Type $i$ adjust to ($k_B$ is removed) :
 \begin{eqnarray}
\lefteqn{S_i=N_i \ln V+\frac{3}{2}N_i \ln \bigg(\frac{2\pi e m_i k_B T}{h^2}\bigg){}}
\nonumber\\
&&{}=N_i\Bigg[\ln V T^{\frac{3}{2}}+\ln\bigg(\frac{2\pi e m_i k_B T}{h^2}\bigg)^{\frac{3}{2}}\Bigg]
\end{eqnarray}
Observation formula (26), where $V$,$T$ are continuous variables, can be treated as formula (6), which is regarded as the result of calculating the continuous entropy $\ln V T^{\frac{3}{2}}$ of a gas molecule, and $\ln\bigg(\frac{2\pi e m_i k_B T}{h^2}\bigg)^{\frac{3}{2}}$ is regarded as an quantized step size denoted by $\Delta_i$.

Go back to the Gibbs paradox, if one still use the conditions $V_1=V_2=V/2,\qquad N_1=N_2=N/2$ , before mixing, the entropy of the two gases is:
\begin{equation}  
\left. \begin{array}{ll}
H_1=\frac{N}{2}\left[\ln\left(\frac{V}{2}T^{\frac{3}{2}}\right)-\ln \Delta_1\right]-\ln((N/2)!)\\
H_2=\frac{N}{2}\left[\ln\left(\frac{V}{2}T^{\frac{3}{2}}\right)-\ln \Delta_2\right]-\ln((N/2)!)\\
\end{array} \right.
\end{equation}
Now, using a new quantized step size $\Delta_s$ , and $\Delta_s<<\Delta_1$,$\Delta_s<<\Delta_2$,so that under the condition with $\Delta_s$, there is no difference between the "symbols" that the two particles can take in the phase space; then, with “non-need for distinguishing” property, the entropy after mixing is:
\begin{equation}H_M=N\left[\ln\left(V T^{\frac{3}{2}}\right)-\ln \Delta_s\right]-\ln(N!)\end{equation}

Combined with the derivation of formula (6), it is easy to know that the increase of entropy brought using  $\Delta_s$ is $-\ln\Delta_i+\ln\Delta_s$. According to the requirements of quantum mechanics, considering the smallest phase cell $h^3$,by using $\Delta_i$, the mixed entropy is:
\begin{eqnarray}
\lefteqn{H_{M\_R}=N\left[\ln\left(V T^{\frac{3}{2}}\right)-\ln \Delta_s\right]-\ln(N!){}}
\nonumber\\
&&{}+2\cdot(N/2)\ln\Delta_s-(N/2)\ln\Delta_1-(N/2)\ln\Delta_2
\nonumber\\
&&{}=N\ln\left(V T^{\frac{3}{2}}\right)-(N/2)\ln\Delta_1-(N/2)\ln\Delta_2-\ln(N!)
\end{eqnarray}

Finally, let formula (29) subtract the two terms of formula (27), that is,:
     \begin{equation}\Delta H=H_{M\_R}-H_1-H_2\approx 0\end{equation}
In other words, when the number of particles is large enough, neither the same particles nor the different particles will cause the entropy change of the order of magnitude $k_B N ln2$.

For an ideal gas, in phase space, the system uses a phase cell to describe its state, and the phase cell itself takes volume and momentum as parameters, which cannot reflect the different kinds of particles. The difference of gas molecular mass only causes the division of phase cell to project to the "space" of volume and temperature. This difference is only related to the number of different molecules, but not to their spatial positions.

\section{Conclusion}

So far, we have a completely different result from the long-standing Gibbs Paradox:  When the number of particles is large enough for an ideal gas, whether or not the gas is the same or different, the diffusion behavior of gas molecules does not cause the change of energy distribution, so it does not cause thermodynamic entropy change.

On the other hand, no matter the same or different kinds of particles, the mixing entropy increases because the particles have more running space. However, this entropy increase is very weak; in a volume space with a huge number of particles, the increase can not be reflected at all. Hence, the debate about the extension of entropy raised by the Gibbs Paradox\cite{1992Jaynes} is greatly weakened, as inserting the board again into the system requires work will, of course, weakly reduce the entropy of the gas molecule.

This also explains why we never have an experiment to determine the mixing entropy corresponding to a gas case which is close to the case as an ideal gas model. Because now our vacuum extraction technology is about $10^{-9}Pa$, in a liter, there's still about $10^9$ particles-a huge number.

The scholars as Jaynes \cite{1992Jaynes},\cite{2011Dieks} believe that the mixing entropy increase of "different" particles is caused by a "different" positions of the “different” particles. This can certainly be seen as an increase in information entropy; however, this introduction has no direct correspondence with the energy of the system, so it should not be treated simply as thermodynamic entropy or statistical entropy. Hence, adiscussionon the relationship between information and energy is helpful to understand and explain a series of new experiments on information and energy in disequilibrium state.\cite{2010Experiments}

\begin{acknowledgements}

In the process of writing this paper, there is a lot of discussion with the collaborator, Chen Xi, colleagues Zhang Jiang and Yao Yao. The emphasis on the difference between energy and information is Chen Xi's suggestion; Section 2.4 is written for Zhang Jiang and Yao Yao. 

Thank for your encourage, Ji Yang from Inst.Semiconductor,CAS! Because the mixing entropy of Gibbs is one of the basic concepts of modern chemistry, the conclusion of this paper makes me fear that there are loopholes and creates the high pressure to me. After all, the progress of science will face the music; new concepts will not be born like Pan An, the jade tree in the wind.

Thanks for the helpful discussions and invitations from Prof Liu Quanhui, Hunan Univ.; thanks for the encourage and discussion from Prof. Wu Jianyong from Georgetown Univ. and Prof. Qian Hong from Univ. of Washington.

After finishing the version of the manuscript in March, 2021,I was told the work of A. Ben-Naim \cite{A.Ben-Naim1}\cite{A.Ben-Naim2} by Prof.Qian Hong.The work shares many same spirits and inspired me although our approaches to the problem are different.

Thanks for the advice in English from Prof. Li Ning from Eastern Washington Univ. and the modification on this paper from Xu Peiran, a postgraduate student from UPenn. 

\end{acknowledgements}

\clearpage

\end{document}